\documentclass{elsart1p}
\usepackage{graphicx}
\usepackage{amssymb}

\newcommand{\mbf}[1]{\mathbf{#1}}

\begin{document}
\begin{frontmatter}
%
%
%
\title{Light-Front Holography \\ and QCD Hadronization at the Amplitude Level\thanksref{label1}}
\thanks[label1]{This research was supported by the Department
of Energy  contract DE--AC02--76SF00515.  SLAC-PUB-13504.}
%
%
\author{Stanley J. Brodsky}
\address{SLAC National Accelerator Laboratory, Stanford University, Stanford, CA 94309, USA}
\ead{sjbth@slac.stanford.edu}
\author{Guy F. de T\'eramond}
\address{Universidad de Costa Rica, San Jos\'e, Costa Rica}
\ead{gdt@asterix.crnet.cr}

\begin{abstract}
Light-front holography allows hadronic amplitudes in the
AdS/QCD fifth dimension to be mapped to frame-independent light-front
wavefunctions of hadrons in physical space-time, thus providing a
relativistic description of hadrons at the amplitude level.  The 
AdS coordinate $z$ is identified with an invariant light-front
coordinate $\zeta$ which separates the dynamics of quark and gluon binding from 
the kinematics of constituent spin and internal orbital angular momentum. The result is a single-variable
light-front Schr\"odinger equation for QCD which determines the eigenspectrum and the light-front wavefunctions of hadrons for general spin and orbital angular momentum.  A new method for computing
the hadronization of quark and gluon jets at the amplitude level using  AdS/QCD light-front wavefunctions is outlined. 
\end{abstract}
\end{frontmatter}
%
\section{Introduction}
The AdS/CFT correspondence~\cite{Maldacena:1997re} between string
states in anti--de Sitter (AdS) space and conformal field theories in physical space-time, modified for color confinement,
has led to semiclassical models for strongly-coupled QCD which provide analytical insight
into its inherently non-perturbative nature, as well as  predictions for  hadronic spectra, decay constants, form factors and wavefunctions.

We have recently shown~\cite{Brodsky:2006uqa,Brodsky:2007hb} that there is a remarkable correspondence between the AdS description of hadrons and the Hamiltonian formulation of QCD in physical space-time quantized on the light front.  A key feature is ``light-front holography''  which allows one to precisely map the $\rm{AdS}_5$ solutions $\Phi(z)$ for hadronic amplitudes  in the fifth dimensional variable $z$ to light-front hadron wavefunctions $\psi_{n/H}(\zeta)$  in physical
space-time evaluated at fixed light-front time $\tau= t + z/c$~\cite{Brodsky:2006uqa,Brodsky:2007hb}.  For two particles $\zeta^2 = \mbf{b}^2_\perp x(1-x),$ which is conjugate to the invariant mass squared ${\cal M}^2  = {\mbf{k}^2_\perp}/{x(1-x)}.$  Light-front holography thus allows hadronic amplitudes in the
AdS/QCD fifth dimension to be mapped to amplitudes in physical space-time. One can derive this connection by showing that one
obtains the identical holographic mapping for matrix elements
of the electromagnetic current and the energy-momentum tensor~\cite{Brodsky:2008pf}.   The mathematical consistency of light-front holography for both the electromagnetic and gravitational~\cite{Brodsky:2008pf} hadronic transition matrix elements demonstrates that the mapping between the AdS holographic  variable $z$ and the light-front variable $\zeta,$ is a general principle. 

Some of the important features of light-front AdS/QCD include:
(a) Effective frame-independent single-particle Schr\"odinger and Dirac equations for meson and baryon wave equations in both $z$ and $\zeta. $ 
For example, the meson eigenvalue equation is 
\begin{equation}
\big[-{d^2\over dz^2} - {1-4 L^2\over 4 z^2}+U(z) \big]\phi(z) = {\cal M}^2\phi(z),
\end{equation}
where the soft-wall potential has the form of a harmonic oscillator  
$ U(z) = \kappa^4 z^2 + 2 \kappa^2(L+S-1).$
(b) The mass spectra formula for mesons  at zero quark mass in the soft-wall model is 
${\cal M}^2 = 4 \kappa^2 (n + L +S/2),$
which agrees with conventional Regge phenomenology. As in the Nambu string model based on a rotating flux tube, the Regge slope is the same for both the principal quantum number $n$ and the orbital angular momentum $L$.  The AdS/QCD correspondence thus builds in a remarkable connection between the string mass $\mu$ in the string  theory underlying an effective gravity theory in the fifth dimension with the orbital angular momentum of hadrons in physical space-time.
(c) The pion is massless at zero quark mass in agreement with general arguments based on chiral symmetry.
(d) The predicted form factors for the pion and nucleons agree well with experiment.  
The nucleon LFWFs  have both $S-$ and $P-$ wave components, allowing one to compute both Dirac and Pauli form factors.
In general the AdS/QCD form factors fall off in momentum transfer squared $q^2$ with a leading power  predicted by dimensional counting rules and the leading twist of the hadron's dominant interpolating operator at short distances. 
Under conformal transformations the interpolating operators transform according to their twist, and consequently the AdS isometries map the twist scaling dimensions into the AdS 
modes~\cite{Brodsky:2003px}. Light-front holography thus provides a
simple semiclassical approximation to QCD which has both constituent counting
rule behavior~\cite{Brodsky:1973kr,Matveev:1973ra} at short distances and confinement at large 
distances~\cite{Brodsky:2008pg,Polchinski:2001tt}.
(e) The timelike form factors of hadrons exhibit poles in the $J^{PC}= 1^{--}$ vector meson channels, an analytic feature  which arises from  the dressed electromagnetic current in AdS/QCD~\cite{Brodsky:2007hb,Grigoryan:2007my}.
(f) The form of the nonperturbative pion distribution amplitude $ \phi_\pi(x)$ obtained from integrating the $ q \bar q$ valence LFWF $\psi(x, \mbf{k}_\perp)$  over $\mbf{k}_\perp$,
has a quite different $x$-behavior than the
asymptotic distribution amplitude predicted from the ERBL PQCD
evolution~\cite{Lepage:1979zb,Efremov:1979qk} of the pion distribution amplitude.
The AdS prediction
$ \phi_\pi(x)  = \sqrt{3}  f_\pi \sqrt{x(1-x)}$ has a broader distribution than expected from solving the ERBL evolution equation in perturbative QCD.
This observation appears to be consistent with the results of the Fermilab diffractive dijet 
experiment~\cite{Aitala:2000hb}, the moments obtained from lattice QCD~\cite{Brodsky:2008pg}, and pion form factor data~\cite{Choi:2006ha}.

\section{Hadronization at the Amplitude Level}

\begin{figure}[!]
 \begin{center}
\includegraphics[width=8.0cm]{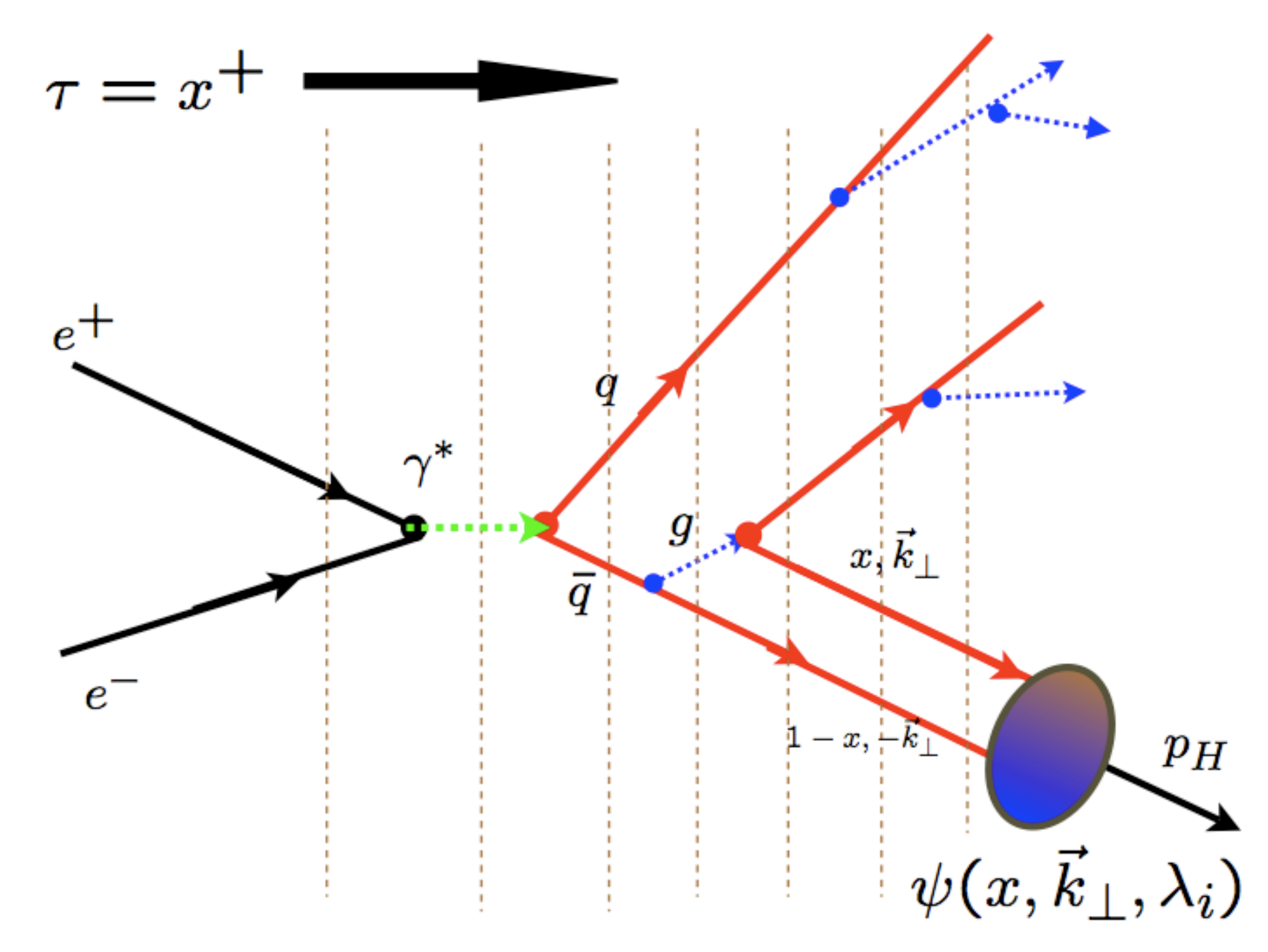}
\end{center}
  \caption{Illustration of an event amplitude generator for $e^+ e^- \to \gamma^* \to X$ for 
  hadronization processes at the amplitude level. Capture occurs if
  $\zeta^2 = x(1-x) \mbf{b}_\perp^2 > 1/ \Lambda_{\rm QCD}^2$
   in the AdS/QCD hard-wall model of confinement;  i.e., if
  $\mathcal{M}^2 = \mbf{k}_\perp^2/x(1-x) < \Lambda_{\rm QCD}^2$.}
\label{fig1}  
\end{figure}

The conversion of quark and gluon partons is usually discussed in terms  of on-shell hard-scattering cross sections convoluted with {\it ad hoc} probability distributions. 
The LF Hamiltonian formulation of quantum field theory provides a natural formalism to compute 
hadronization at the amplitude level~\cite{Brodsky:2008tk}.  In this case one uses light-front time-ordered perturbation theory for the QCD light-front Hamiltonian to generate the off-shell  quark and gluon T-matrix helicity amplitude  using the LF generalization of the Lippmann-Schwinger formalism:
\begin{equation}
T ^{LF}= 
{H^{LF}_I } + 
{H^{LF}_I }{1 \over {\cal M}^2_{\rm Initial} - {\cal M}^2_{\rm intermediate} + i \epsilon} {H^{LF}_I }  
+ \cdots 
\end{equation}
Here   ${\cal M}^2_{\rm intermediate}  = \sum^N_{i=1} {(\mbf{k}^2_{\perp i} + m^2_i )/x_i}$ is the invariant mass squared of the intermediate state and ${H^{LF}_I }$ is the set of interactions of the QCD LF Hamiltonian in the ghost-free light-cone gauge~\cite{Brodsky:1997de}.
The $T^{LF}$ matrix elements are evaluated between the out and in eigenstates of $H^{QCD}_{LF}$. The LFWFs of AdS/QCD can be used as the interpolating amplitudes between the off-shell quark and gluons and the bound-state hadrons.
Specifically,
if at any stage a set of  color-singlet partons has  light-front kinetic energy 
$\sum_i {\mbf{k}^2_{\perp i}/ x_i} < \Lambda^2_{\rm QCD}$, then one coalesces the virtual partons into a hadron state using the AdS/QCD LFWFs.   This provides a specific scheme for determining the factorization scale which  matches perturbative and nonperturbative physics.
The event amplitude generator is illustrated for $e^+ e^- \to \gamma^* \to X$ in Fig. \ref{fig1}.

This scheme has a number of  important computational advantages:

(a) Since propagation in LF Hamiltonian theory only proceeds as $\tau$ increases, all particles  propagate as forward-moving partons with $k^+_i \ge 0$.  There are thus relatively few contributing
 $\tau-$ordered diagrams.

(b) The computer implementation can be highly efficient: an amplitude of order $g^n$ for a given process only needs to be computed once.  In fact, each non-interacting cluster within $T^{LF}$ has a numerator which is process independent; only the LF denominators depend on the context of the process.

(c) Each amplitude can be renormalized using the ``alternate denominator'' counterterm method~\cite{Brodsky:1973kb}, rendering all amplitudes UV finite.

(d) The renormalization scale in a given renormalization scheme  can be determined for each skeleton graph even if there are multiple physical scales.

(e) The $T^{LF}$ matrix computation allows for the effects of initial and final state interactions of the active and spectator partons. This allows for novel leading-twist phenomena such as diffractive DIS, the Sivers spin asymmetry and the breakdown of the PQCD Lam-Tung relation in Drell-Yan processes.

(f)  ERBL and DGLAP evolution are naturally incorporated, including the quenching of  DGLAP evolution  at large $x_i$ where the partons are far off-shell.

(g) Color confinement can be incorporated at every stage by limiting the maximum wavelength of the propagating quark and gluons.

This method retains the quantum mechanical information in hadronic production amplitudes which underlie Bose-Einstein correlations and other aspects of the spin-statistics theorem.
Thus Einstein-Podolsky-Rosen quantum theory correlations are maintained, even between far-separated hadrons and  clusters ~\cite{Abelev:2008ew}.
A similar off-shell T-matrix approach was used to predict antihydrogen formation from virtual positron--antiproton states produced in $\bar p A$ 
collisions~\cite{Munger:1993kq}.


%
%
%

%
\end{document}